\documentclass[twocolumn,showpacs,floatfix,preprintnumbers,amsmath,amssymb,pra]{revtex4}

\usepackage{epsfig}
\usepackage{graphicx}
\usepackage{dcolumn}
\usepackage{bm}

\begin{document}

\title{One- and two-photon ionization cross sections of the laser-excited $6s6p\,^1P_1$ state of barium}
\author{John R. Tolsma}
\email{john.tolsma@colorado.edu}
\affiliation{Department of Physics and JILA, University of Colorado, Boulder, CO 80309-0440, USA}
\author{Daniel J. Haxton}
\affiliation{Department of Physics and JILA, University of Colorado, Boulder, CO 80309-0440, USA}
\author{Rekishu Yamazaki}
\affiliation{School of Electrical and
Computer Engineering and Department of Physics, Purdue University,
West Lafayette, Indiana 47907}
\author{Daniel S. Elliott}
\affiliation{School of Electrical and
Computer Engineering and Department of Physics, Purdue University,
West Lafayette, Indiana 47907}
\author{Chris H. Greene}
\affiliation{Department of Physics and JILA, University of Colorado, Boulder, CO 80309-0440, USA}

\date{\today}

\begin{abstract}
Stimulated by a recent measurement of coherent control in photoionization of atomic barium, we have calculated one- and
two-photon ionization cross sections of the aligned $6s6p\,^1P_1$ state of barium in the energy range between the
$5d_{3/2}$ and $5d_{5/2}$ states of Ba$^+$.  We have also measured these photoionization spectra in the same energy
region, driving the one- or two-photon processes with the second or first harmonic of a tunable dye laser, respectively.
Our calculations employ the eigenchannel R-matrix method and multichannel quantum defect theory to calculate the
rich array of autoionizing resonances in this energy range. The non-resonant two-photon process is described using
lowest-order perturbation theory for the photon-atom interactions, with a discretized intermediate state one-electron continuum.  The calculations
provide an absolute normalization for the experiment, and they accurately reproduce the rich resonance structures in both the one and two-photon
cross sections, and confirm other aspects of the experimental observations.
These results demonstrate
the ability of these computationally inexpensive methods to reproduce experimental observables in one- and two-photon
ionization of heavy alkaline earths, and they lay the groundwork for future studies of the phase-controlled interference between one-photon and two-photon ionization processes.

\end{abstract}

\pacs{32.80.Fb,           
32.80.Qk,     
33.80.Rv,           
32.80.Rm           
}

\keywords{phase lag, asymmetric photoelectron angular distribution, PAD, barium, autoionizing resonances}
\maketitle
\section{Introduction}
Through two-pathway coherent control of optical interactions, the outcome of a laser-driven interaction in an atom or
molecule can be controlled by varying only the optical phase difference between various coherent laser
fields~\cite{BrumerS86a, BrumerS86b,ChenYE90}.  Mutual phase coherence between the various optical fields and careful
matching of optical wavefronts is required, but upon satisfaction of these conditions, various groups have used coherent
control to modulate excitation rates of transitions~\cite{ChenYE90,ParkLG91}, branching ratios of
photodissociation~\cite{ZhuKLLTG95,ShnitmanSGYSCB96}, and even directional photocurrents in unbiased
semiconductors~\cite{hache97a, hache97b}.  When two or more dissociation channels are excited via interfering interactions,
coherent control can be used to control the branching ratio into these different channels.  Control is strongest when the
transition amplitudes for the two pathways are of equal magnitude, and when the sinusoidal modulation of one product
state and sinusoidal modulation of the other product state are $\pi$ out of phase with one another.  The latter
condition leads to constructive interference for one product state under the conditions that yield destructive
interference for the other.  Thus, the observation by Gordon \emph{et al.}~\cite{ZhuKLLTG95,ZhuSFWSG97} of a large
phase difference between product states when photodissociating molecular $HI \rightarrow HI^+$ and $H + I^+$ was an
important advance in the field.

The origin of the phase difference between product states generated significant
interest~\cite{ZhuSFWSG97,KhachatrianBZGS02,FissZGS99,LambropoulosN99,LefebvreBrion97,NakajimaZL97b,LyrasB99,Lee98,Seideman98,Seideman99,ApalateguiSL01,YamazakiE07a, YamazakiE07b},
and ultimately channel mixing in the vicinity of discrete states embedded in the continuum (autoionizing or
predissociating states) was identified as the dominant mechanism~\cite{ZhuSFWSG97,FissZGS99,LambropoulosN99}.
It is well known from the scattering theory formulations, e.g. by Feshbach~\cite{feshbach} and by Fano~\cite{fano61}, that the wave function components of the resonance
and of the background can interfere quantum mechanically to alter the associated scattering observables,
such as differential cross sections, alignment or orientation, or other quantities dependent on the phase of the scattered particles.  Such interference occurs only when the operators labeling the wavefunction components fail to commute with the observable of interest. The large phase-dependent variation in these observables is well- suited for the quantum control of the scattering, as well as other important applications such as the generation of
cold molecules and degenerate quantum gases.

While there is qualitative support for the primary role that autoionizing or predissociating resonances play in modulating the
phase difference between alternative pathways in coherent control experiments, there are no cases in which theory and experiment agree
quantitatively, presumably because of the complexity of molecular systems for which these experiments and calculations have
been carried out.  For this reason, Yamazaki and Elliott~\cite{YamazakiE07a, YamazakiE07b} recently measured the phase
shift in atomic barium, using the interference between one- and two-photon ionization of the aligned $6s6p\,^1P_1$ state
of barium in the energy range between the $5d_{3/2}$ and $5d_{5/2}$ states of Ba$^+$.  Because the final states reached via the one- and two-photon processes are of
opposite parity, the interference manifests itself as a modulation of the angular distribution of the
photoelectrons, while the total cross sections are unaffected. In contrast to the usual photoelectron angular distribution experiments carried out where only a fixed number of electric dipole photons excite the observed final photoelectron energy~\cite{FanoDill72,Lindsay92}, the angular distribution is modulated in the present case asymmetrically with regard to parity.~\cite{BrumerS86a, BrumerS86b,ChenYE90, Zeldovich92} Observations of this modulation
showed strong variations in the phase difference near several of the autoionizing
resonances, with the largest phase shift of nearly $2\pi$ near the
$5d \: ^2D_{5/2} 15f_{3/2}$
two-photon peak at 47,216.5 cm$^{-1}$.

A calculation of these phase variations in the coherent control process requires an accurate description of the
outgoing electronic waves, including their magnitudes and phases for various orbital angular momenta and spin orientations.  This work reports
a joint theoretical and experimental study of the cross sections for one- and two-photon ionization of
atomic barium from the aligned $6s6p\,^1P_1$ state.  Determining these separate cross sections is a key prerequisite to an eventual description of the more sensitive phase interference effect.
Following a discussion of the experimental measurements in Sec.~\ref{sec:experiment}, we summarize our
theoretical approach for calculating the spectra in Sec.~\ref{sec:theory}.
Sec.~\ref{sec:spectra} compares the measured spectra with theory.

\section{Experiment}\label{sec:experiment}
\begin{figure}
\includegraphics[width=8cm]{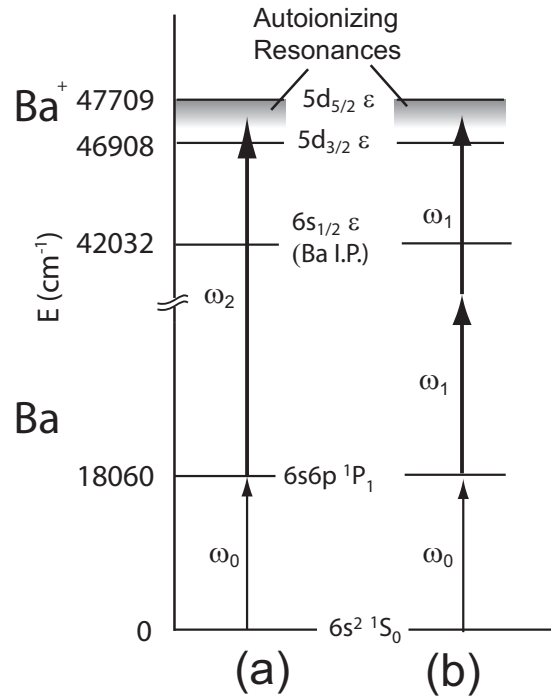}\\
  \caption{An atomic barium energy level diagram, showing the relevant energy levels and excitation schemes in
our work.  (a) The scheme for exciting the $6s6p\,^1P_1$ intermediate state to the even parity autoionizing
resonances through the absorption of a single UV photon of wavelength $\lambda_2$ $\sim$ 343 nm. (b) The scheme
for accessing the odd continuum at the same final state energy through two-photon absorption ($\lambda_1$ $\sim$ 685 nm) from the same intermediate
state is shown.}\label{fig:AIRenergy}
\end{figure}

Fig.~\ref{fig:AIRenergy} presents an energy level diagram with the relevant atomic states of barium and the experimental excitation schemes.
Fig.~\ref{fig:AIRenergy}(a) shows the pathway leading to excitation of the even autoionizing resonances,
while Fig.~\ref{fig:AIRenergy}(b) displays the pathway to the odd-parity autoionizing resonances.  In each case, the experiment first promotes
ground state ($6s^2\,^1S_0$) barium to the $6s6p\,^1P_1$ intermediate state by a linearly-polarized laser beam at frequency
$\omega_0$, tuned near the resonance wavelength of approximately 553.7 nm.  The subsequent absorption of two photons
of the output of a second tunable dye laser at frequency $\omega_1$ (wavelength $\lambda_1$ $\sim$ 685 nm), or of a single
photon from the second harmonic of this field $\omega_2$ (wavelength $\lambda_2$ $\sim$ 343 nm) ionizes the atom to a
continuum state between the $5d_{3/2}$ and $5d_{5/2}$ thresholds, 46,908.76 and 47,709.72 cm$^{-1}$ above the atomic ground
state, respectively~\cite{KarlssonL99}.

Because the excitation energy of the interaction reaches a final state energy that exceeds the energy of the first excited ($5d_{3/2}$) state of Ba$^+$, two states
of the residual Ba$^+$ core can be populated.  We distinguish these through measurements of the kinetic energies of the
corresponding photoelectrons.  Upon excitation of the $6s_{1/2}$ continuum channel, the photoelectron energy is
approximately 0.6 eV, while the
$5d_{3/2}$ channel correlates with the 0-0.1 eV electrons.  We resolve the photoelectron energies using a microchannel
plate detector assembly, described below.  This detector also allows us to measure the photoelectron angular
distributions, as described in Refs.~\cite{YamazakiE07a,YamazakiE07b}.

\begin{figure}
  \includegraphics[width=8cm]{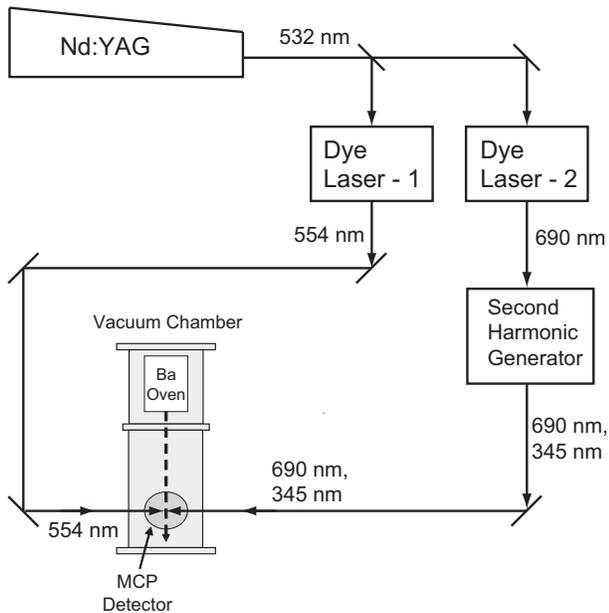}\\
  \caption{The experimental setup.  A Q-switched Nd:YAG laser pumps two tunable dye lasers, producing beams of
wavelength 554 nm and $\sim$685 nm. The former prepares Ba in the $6s6p\,^1P_1$ state, while the latter ionizes
this state.  The second harmonic generator produces frequency-doubled light in the range 337 nm - 347 nm for
single-photon ionization of the $6s6p\,^1P_1$ state.  The second harmonic generator is omitted for measurement of
the two-photon ionization spectra.  The vacuum chamber contains the barium oven and microchannel
plate (MCP) detector.}\label{fig:AIRsetup}
\end{figure}

A schematic of the experimental setup is depicted in Fig.~\ref{fig:AIRsetup}.
The second harmonic output (532 nm) of a Q-switched Nd:YAG laser (10 Hz repetition rate) is used to pump two dye
lasers, a homemade Littman-type laser (Dye Laser - 1) with one longitudinally-pumped amplifier stage, and a Spectra
Physics PDL-2 (Dye Laser - 2).  We use Rhodamine 590 dye in methanol in the first laser to generate the beam at
frequency $\omega_0$ (corresponding to a wavelength of 553.7 nm) and pulse energy 60 $\mu$J.  This laser is kept at a set wavelength during the experiment.
To generate the laser beam at frequency $\omega_1$, whose wavelength we vary in the range from 674-694 nm, we use
LDS 698 dissolved in methanol in the PDL-2 laser.  Scanning the wavelength of this laser allows us to tune across
the continuum in the energy range between the $5d_{3/2}$ and $5d_{5/2}$ thresholds. For the ultraviolet (UV)
generation, the output of the PDL-2 is sent into a Type-I BBO nonlinear crystal, mounted on a motorized rotation
stage controlled with the phase-matching servo system (Inrad Autotracker).  The output of both lasers is in nearly
a single longitudinal mode, and we set each of the laser polarizations to be horizontal before sending these beams
into a vacuum chamber for the interaction with the atom beam.  The pulse duration of each laser output is about 6 ns
for $\omega_0$ and 11 ns for $\omega_1$. The wavelength of the PDL-2 output is controlled by a stepping motor
attached to a grating inside the oscillator cavity, with a minimum step size of 0.0030 nm @ 690 nm
(0.063 cm$^{-1}$).  During a scan, we monitor the wavelength of the output of
PDL-2 using a wavemeter with a specified frequency accuracy of 0.02
cm$^{-1}$ (Burleigh WA-4550).  In practice, however, we observed inconsistencies in the wavelength readings of the wavemeter, which we attribute to degradation in the coatings of the wavemeter etalons, and we assign an uncertainty of 0.1 nm to the absolute wavelength reading  (2 cm$^{-1}$ in frequency).  As we discuss later, we choose our absolute frequency calibration for the red two-photon spectrum as well as the UV one-photon spectrum using the known energy of the $6s7p$ $^1$P$_1$ bound state.

The first excitation laser pulse
$\omega_0$ and the second excitation laser pulse (either at frequency $\omega_1$ or $\omega_2$) are sent into a vacuum chamber from opposite sides.
The intensity peaks of the first and second excitation pulses coincide at the interaction region; the relative timing of the pulses is adjusted using a fast photodiode (with a resolution of
0.5 ns) and an optical delay.  The pulse energy at $\omega_1$ ($\omega_2$) is 1.5 mJ (14 $\mu$J).  We weakly focus
each of the laser beams into the interaction region to a beam radius of  $w$=300 $\mu$m, where $w$ is the radius in
which the intensity decreases by a factor
$e^{-2}$ from that of the peak value.

During a scan, we step the wavelength of the $\omega_1$-beam with an increment of 0.012 nm.  We limit the wavelength range of each scan to minimize the effect of pulse energy variations.  The variations in the pulse energy of the 690 nm laser pulse are about 15\%, while those of the UV pulses are about 20\%.  The overlap in adjacent scans is always sufficient to include at least one autoionizing resonance that is common to both, allowing us to piece the individual scans together to form the full scans.  We adjust the amplitude of the individual scans slightly (on the order of 10\%) to correct for differences in the laser pulse energy.  The two-photon spectra consist of 4 individual scans, while the UV spectra are composed of 8 individual scans.

The atomic barium beam is generated in an effusive oven inside the
vacuum system.  We maintain the temperature of the oven at 650 $^{\circ}$C during the experiment. The atom beam
generated from the oven is collimated using an aperture positioned
before the interaction region.  The diameter of the atomic
beam at the interaction region is approximately 1 mm, with an estimated atom density of 4.3$\times$10$^7$ cm$^{-3}$.
The background pressure around the interaction region is kept around 1$\times$10$^{-8}$ torr.

\begin{figure}
\resizebox{1.0\columnwidth}{!}{\includegraphics*[0.6in,3.7in][7.8in,7.0in]{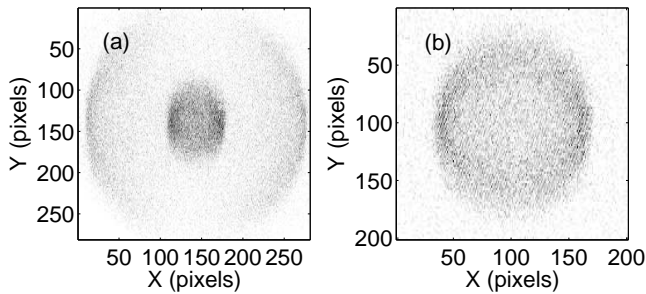}}\\
  \caption{An example of an image of the photoelectrons incident upon the MCP detector.  In (a), the collecting
field is $\sim$11.4 V/cm, and we clearly see the image resolved by kinetic energy.  The outer (inner) ring consists of
photoelectrons that correlate to the $6s_{1/2}$ ($5d_{3/2}$) core state. In (b), the collecting field is reduced
to $\sim$2.9 V/cm in order to expand the $5d_{3/2}$ image. The laser polarization is oriented along the $x$-axis.}\label{fig:PADexample}
\end{figure}

The photoelectrons from the interaction are collected using a microchannel plate (MCP)
detector assembly~\cite{helm}.  The detailed explanation of the detector and the image processing for the product-resolved
imaging can be found elsewhere~\cite{YamazakiE06,YamazakiE07a, YamazakiE07b} and we only discuss it briefly here.  A
pair of flat biased conducting meshes (81\% transmitting) is constructed around the interaction region, which we use to
generate a uniform dc electric field in the interaction region.  The photoelectrons ejected by the barium atoms are
accelerated toward the upper mesh by the bias field ($\sim$10 V/cm).  The photoelectrons are amplified in the MCP, and
each produces a pulse of light on the phosphor screen mounted behind.  We image the phosphorescence from the phosphor
screen using a CCD camera interfaced to a laboratory PC through a framegrabber plug-in card, and accumulate and store
the image in a laboratory PC for a further image processing.  We show an example of an image of the photoelectrons
Fig.~\ref{fig:PADexample}.
The image collected can be resolved in kinetic energy and also provides the photoelectron angular distributions
for each energy-resolved electron channel, as illustrated in the figure.  We obtain the electron yield for each continuum
channel by simply summing the electron counts in the image for the corresponding energy region.  The typical number of
electrons collected per laser pulse is about 10-150 electrons in a CCD region of 380$\times$380 pixels.  At every
wavelength, we accumulate 100 images.

The noise generated from the 554 nm and 685 nm laser pulse is less than 1 count/shot, while the noise from the UV probe,
including the noise caused from the scattering of the UV pulse and the ionization of the background gas inside the
vacuum system was about 13 counts/shot, uniformly distributed across the MCP detector.

We measured the $5d_{3/2}$ threshold energy of 46,890 cm$^{-1}$, differing by 18 cm$^{-1}$ from the value of
Ref.~\cite{KarlssonL99}.  This difference may be due to ac Stark shifts by the laser pulses, field
ionization by the 10 V/cm dc collection field, or false counting of electrons with low kinetic energies.  As the
photoelectron kinetic energy decreases, the
image radius becomes smaller and images of neighboring electrons start to
overlap.  False counting can occur for either extremely large electron
numbers in the image and/or the small kinetic energy of the electrons.


\section{Theory}\label{sec:theory}

We use atomic units throughout this section, unless otherwise specified.

Multichannel quantum defect theory (MQDT)~\cite{fanorau, seaton83} is well-established as an
effective method to calculate photoionization spectra of atoms.  The combination of MQDT and the R-matrix
method~\cite{burkeRmatrix} has been used to calculate rates of photionization and electron scattering processes
for atoms and molecules. In particular, much attention has been
given~\cite{AymarR79,GreeneKim87, GreeneTheo, Aymar90, GreeneAymar91, WGA93a, WGA93b, WGAcoop, WGA94, lecomte, maeda, OrangeRev}
to the problem of calculating the photoionization spectra of alkali earth atoms.
We follow the methodology of these papers.  In particular, we use a two-active-electron
treatment of the barium atom, accounting for the [Xe] core using a model potential~\cite{AymarR79} with spin-orbit terms.
For a complete review of the techniques employed in this work, see Ref.~\cite{OrangeRev}.

The initial states (denoted $\vert \Psi_0 \rangle$)
and intermediate states for the two photon calculation (denoted $\vert \epsilon \rangle$) are obtained from
configuration interaction (CI) calculations, and the final states (denoted $\vert \Psi_f \rangle$) are
calculated using the eigenchannel R-matrix framework~\cite{eigenchannel}.  We employ partial waves
up to $l$=4 in the
single-particle basis.  Our basis includes orbitals that obey a zero boundary condition at the R-matrix
radius, which are used for all states, and orbitals that are nonzero on the boundary,
used only in the final states.  These orbitals are termed ``closed-type'' and ``open-type'' respectively.
Those configurations with no open orbitals are denoted ``closed'' and those with one open orbital are
denoted ``open.''  Configurations are chosen according to their single particle principal quantum numbers.
We select configurations by restricting the highest value of the single particle principal quantum number included
to lie below a specified value, and by restricting the highest value that both single particle principal quantum
numbers may simultaneously exceed to a lower value, in both cases irrespective of $l$.

The pseudopotential we employ~\cite{AymarR79}
does not exactly reproduce the experimental energies.  It is optimized to most accurately
reproduce the ionization energies of Ba$^+$ and performs less well in reproducing the
neutral energies.  In some studies this has been improved by incorporation of a dielectronic polarization term in the Hamiltonian, but this has not been done here.
Instead,
we adjust the energy of the initial state (for both calculations)
and a few of the experimentally probed intermediate states (for the two photon calculation).  We list these
adjustments in Table~\ref{tab1}.  We also fix the channel thresholds, but these
adjustments are smaller: by +0.73, -4.5, and +3.8 cm$^{-1}$ for the $6s_{1/2}$,
$5d_{3/2}$, and $5d_{5/2}$ channels.

\begin{table}
\begin{ruledtabular}
\begin{tabular}{cccc}
State             &  J  & experimental  &  energy adjustment \\
                  &     &  energy       & (experimental --     \\
                  &     &  (cm$^{-1}$)   & calculated)       \\
\hline
$^1$P$_1$ (6s6p) $^*$  &  1  & 18060.261          &  -208.80  \\
$^3$D     (5d7s) $^\dag$ &  1  & 32805.169          &  -34.79   \\
$^3$D     (5d7s) $^{\dag\dag}$ &  2  & 32943.774          &  -20.54 \\
$^1$D     (5d7s) $^{\dag\dag}$ &  2  & 33796.011          &  22.77 \\
$^3$S     (6s8s) $^{\dag\dag}$ &  1  & 33905.358          &  47.92 \\
\end{tabular}
\end{ruledtabular}
\begin{tabular}{l}
 $^*$ Initial state \\
 $^\dag$ Intermediate state resonance visible in two-photon \\
 \qquad \qquad calculation \\
 $^{\dag\dag}$ Other intermediate states used in two-photon calculation \\
\end{tabular}
\caption{Initial and intermediate states whose energies are fixed in the calculations to the
experimental values listed. \label{tab1}}
\end{table}

\subsection{One-photon calculations}

The one-photon calculations precisely follow the methodology of, for instance, Ref.~\cite{WGA93a}.  For these
calculations we employ a R-matrix box size of 28$a_0$.  We employ MQDT to enforce the boundary condition for the
wavefunction in the closed $5d_{5/2}$ channel, and also the higher $6p_{1/2}$ and $6p_{3/2}$ channels.  Thus,
these channels are ``weakly closed'' in MQDT terminology, and the exponentially decaying wavefunctions in these
channels may have a nonzero value at the R-matrix boundary.  Higher channels are strongly closed and obey a zero
boundary condition there.

For the one photon calculation, our configuration list includes
closed configurations in which the lowest principal quantum number is no greater than nine
(e.g., the 6$s$ through the 9$s$), and in which the highest principal quantum number is no greater than
15 and 23 for the initial and final states respectively.  This gives 704 configurations for the initial
state and 594, 1454, and 1998 configurations for final states with J=0, 1, and 2 respectively.

\subsection{Two-photon calculations}

Two-photon atomic ionization has been treated theoretically by several methods.
A recent R-matrix Floquet treatment is described in Ref.~\cite{rmatfloq}.
Some calculations~\cite{twophomag,robgao,Bachau2004,SCG2004, CohenCalcium2phot2006} have utilized the
R-matrix-plus-MQDT-based descriptions of two-photon ionization in light alkaline earths, e.g.,  magnesium and
calcium, as in the present work on the heavier barium atom.
Two-color two-photon ionization of barium was treated using a combination
of the R-matrix method, MQDT, and the Feshbach projection operator formalism~\cite{feshbach}, within an adiabatic
Floquet model, in Ref.~\cite{Lyras2000}.
 An $L^2$ method employing an integral equation for the K-matrix
was used in Ref.~\cite{twophotCI}.

The two-photon ionization amplitude in
second-order perturbation theory for the radiation-matter
interaction can be defined
\begin{equation}
T_{f0}(\omega) = \lim_{\eta\rightarrow 0^{^+}}\int \ d\varepsilon \ \frac{\langle\Psi_{f}\vert\vec{D} \cdot \hat{e}\vert\varepsilon\rangle\langle\varepsilon\vert\vec{D} \cdot \hat{e}\vert\Psi_{0}\rangle}{(E_0 + \omega - \varepsilon -i\eta)} \ .
\label{eq1}
\end{equation}
The integration in this equation is meant to include a summation over bound states and an
integration over continuum states.
When above-threshold ionization (ATI) is energetically forbidden, the outgoing wave boundary condition in
Eq.~(\ref{eq1}) can be replaced by a standing wave boundary condition.  This approximation, combined with
a discretization of the continuum, is the strategy adopted here and in, for example,
Refs.~\cite{CohenCalcium2phot2006,Bachau2004, SCG2004, Lambro95}.  In Ref.~\cite{Lambro95} an extrapolation
proceedure was used to obtain spectra in the ATI energy range; in its most straightforward application,
however, a treatment using discretized standing-wave continuum states is inappropriate when ATI is allowed,
and the Dalgarno-Lewis method~\cite{dlewis} is often applied~\cite{robgao,twophomag}.

We treat the continuum intermediate states
$\vert\varepsilon\rangle$ as box states, using a box radius of 60$a_0$.  We find that this radius is
sufficient to converge the results of the calculation.
The generalized total cross section for a two-photon transition from a pure initial state
can then be represented in terms of T as
\begin{equation}
\sigma^{(2)}(\omega) = 8\pi^3 \alpha^2\omega^2 \sum_{f}\vert T_{f,0}(\omega)\vert^2 \quad ,
\end{equation}
which is defined to have units of (length)$^4$ (time), such that the probability per unit time, $R$, of a transition
is
\begin{equation}
R = {\cal J}^2 \sigma^{(2)} \quad ,
\end{equation}
where ${\cal J}$ in this equation only is the photon number flux, i.e. the number of photons per unit area per unit time.

The two-photon calculations include closed configurations in which the lowest principal quantum number
is no greater than nine, and in which the number of box states per $lj$-orbital is no greater than
21, 21, or 25 for the initial, intermediate, and final states respectively.  This gives us 1270
configurations for the initial J=1 state; 522, 1270, and 1750 configurations for intermediate states with
J=0, 1, and 2, respectively; and 1584, 2160, and 2304 closed configurations for final states with J=1, 2, and 3,
respectively.

The larger box radius utilized in the two-photon calculations necessitates that the $5d_{5/2}$
channel is the only one treated as weakly-closed, and in contrast to the one-photon calculations we enforce a zero boundary
condition at the R-matrix boundary for the higher $6p_{1/2}$ and $6p_{3/2}$ channels.

\subsection{Hyperfine depolarization}

As described in Ref.~\cite{WGA93a}, it is necessary to include the effect of hyperfine depolarization of the
initially excited $^1$P$_1$ state.  This state is initially excited to the $m_J = 0$ sublevel, but the presence
of nuclear spin and the associated hyperfine splittings will in general cause the electronic angular momentum
to precess about the total angular momentum, populating the $m_J = \pm 1$ sublevels, which process is called
hyperfine depolarization~\cite{FanoMacek73RMP, GreeneZare1982}.  Isotopes of barium with nonzero nuclear spin
occur with 18\% natural abundance
and are therefore subject to hyperfine depolarization.

In Ref.~\cite{WGA93a} a parameter $g$ describes the amount of hyperfine depolarization.
A value of $g=1.0$ corresponds to no depolarization; the maximum amount of hyperfine depolarization, given
a pure initial excitation of the $m_J=0$ sublevel, corresponds to a value of $g=0.896$.  We find, however, that
the best agreement with experiment is obtained with a value of $g \approx 0.82$.  This fact indicates that
the initial excitation of the $^1$P$_1$ state might initially create some population in the $m_J = \pm 1$ sublevels, which in
turn could suggest that this transition may be partially saturated in the experiment.  This conclusion is consistent with the conditions of the experiment.  (The pulse energy, pulse duration and a beam size of $\sim$0.5 mm yield a pulse area of several thousand.  Thus a perpendicular field component of the $\omega_0$-laser whose relative amplitude was even 10$^{-3}$ would be sufficient to promote a non-negligible population in the $m_J = \pm 1$ sublevels.)

\section{Spectra}\label{sec:spectra}

%
%
%

Figs.~\ref{fig:AIRmany1s}, \ref{fig:AIRmany1d}, \ref{fig:AIRmany2s} and \ref{fig:AIRmany2d} show the product-resolved photoionization partial cross sections obtained, in the
energy region between the $5d_{3/2}$ and $5d_{5/2}$ thresholds, for
one-photon and two-photon excitation from the $6s6p\,^1P_1$ state.  
All the spectra
show many final state autoionizing peaks converging to the $5d_{5/2}$ threshold, modulating a continuum of
slowly varying background ionization.

The abscissae in Figs.~\ref{fig:AIRmany1s}, \ref{fig:AIRmany1d}, \ref{fig:AIRmany2s} and \ref{fig:AIRmany2d} are labeled in terms of
the energy above the initial $^1$P$_1$ state, which lies at 18060.261 cm$^{-1}$ above the ground
state of the neutral.
The $5d_{3/2}$ and
$5d_{5/2}$ thresholds are located at red wavelengths of 693.277
nm and 674.549 nm (whose doubled wavenumbers are at 28848.50 and 29649.46 cm$^{-1}$),
calculated from the ionization threshold energy of
46908.76 cm$^{-1}$ and 47709.72 cm$^{-1}$, respectively, and the $6s6p\,^1P_1$
state energy \cite{KarlssonL99}.  The threshold for
$5d_{3/2}$ can be observed as a sudden drop of the signal, near a
red wavelength of 694 nm (28818 cm$^{-1}$), in the one and two-photon
slow electron spectra.

For both spectra, partial cross sections into the slow electron channels are much
smaller than for the faster electron channels.  In the one-photon spectra,
there are many resonance structures with relatively large background ionization;
in contrast, the two-photon spectra show more pronounced autoionizing peaks with
simpler structures.

Because the experimental data are not absolutely normalized, we have normalized
the experiment to theory.  In doing so we use the same constant
for the fast ($6s_{1/2}$) and slow ($5d_{3/2}$) partial cross sections.

\subsection{One-photon cross sections}

\begin{figure*}
\begin{center}
\begin{tabular}{c}
\resizebox{2.0\columnwidth}{!}{\includegraphics{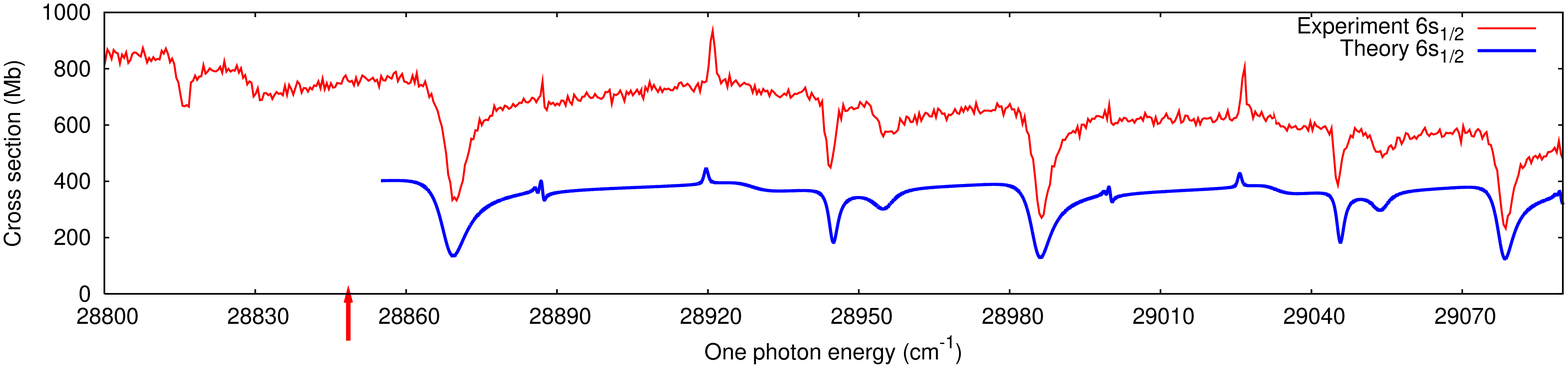}} \\
\resizebox{2.0\columnwidth}{!}{\includegraphics{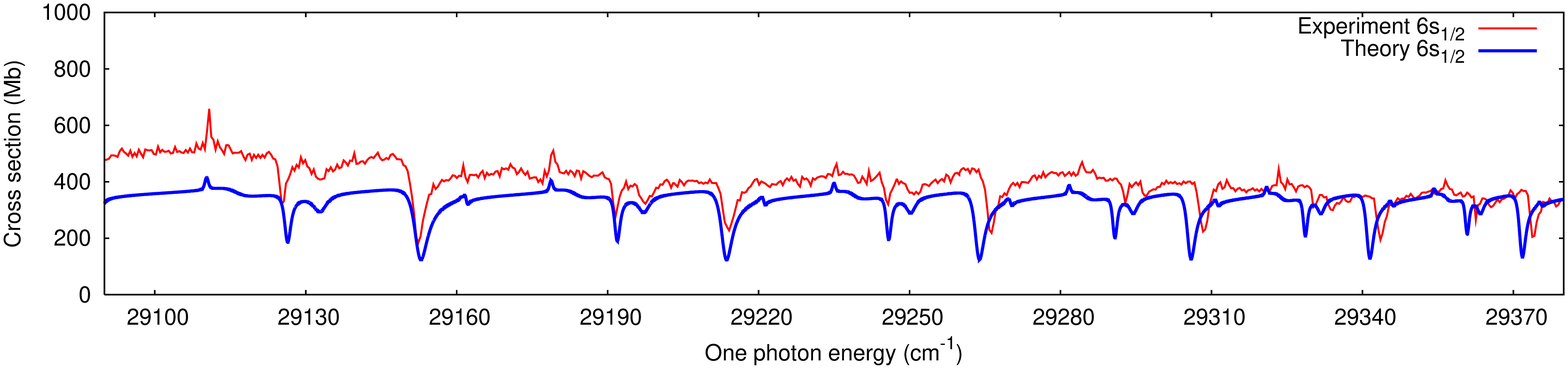}} \\
\resizebox{2.0\columnwidth}{!}{\includegraphics{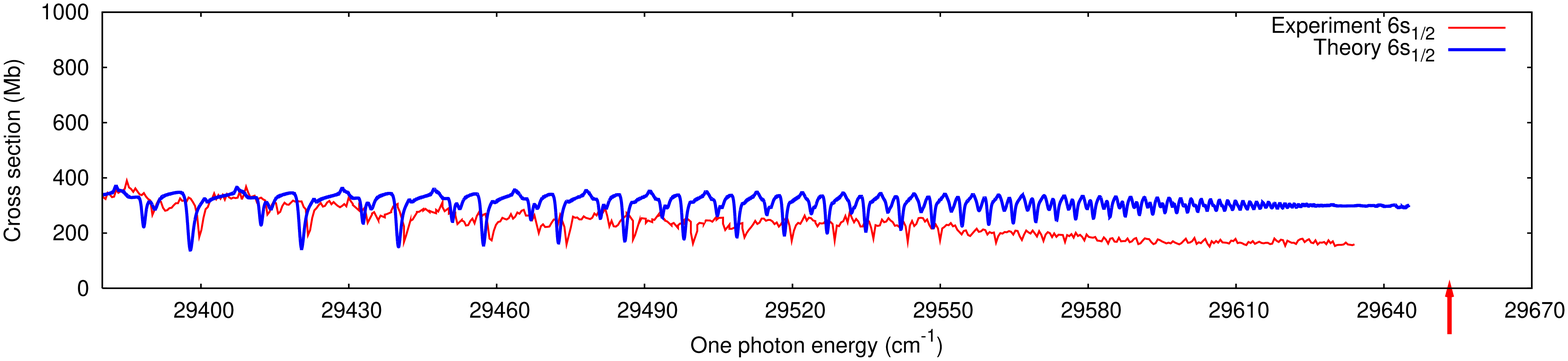}} \\
\end{tabular}
\end{center}
\caption{One photon cross sections: $6s_{1/2}$. \label{fig:AIRmany1s}}
\end{figure*}

\begin{figure*}
\begin{center}
\begin{tabular}{c}
\resizebox{2.0\columnwidth}{!}{\includegraphics{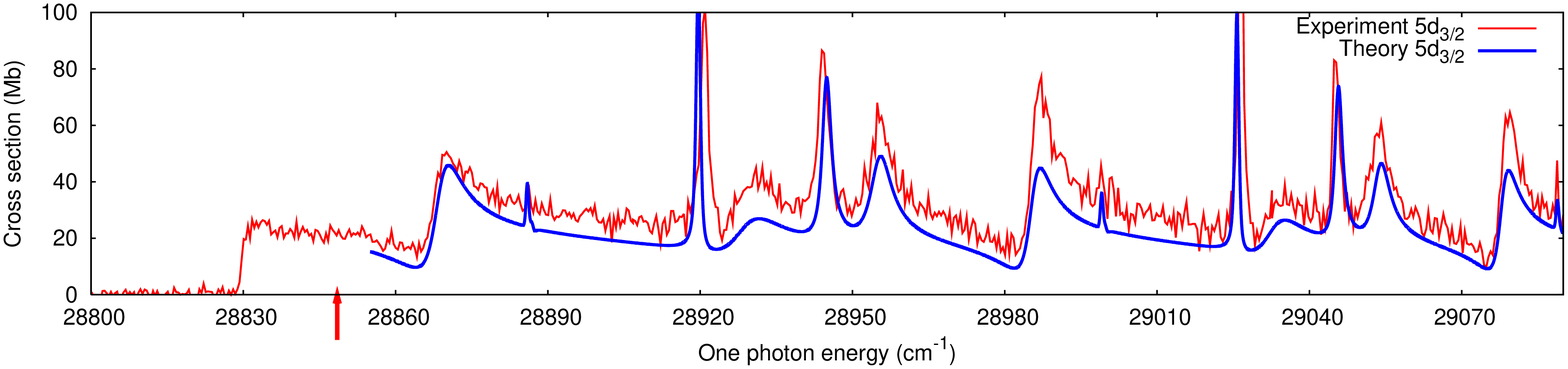}} \\
\resizebox{2.0\columnwidth}{!}{\includegraphics{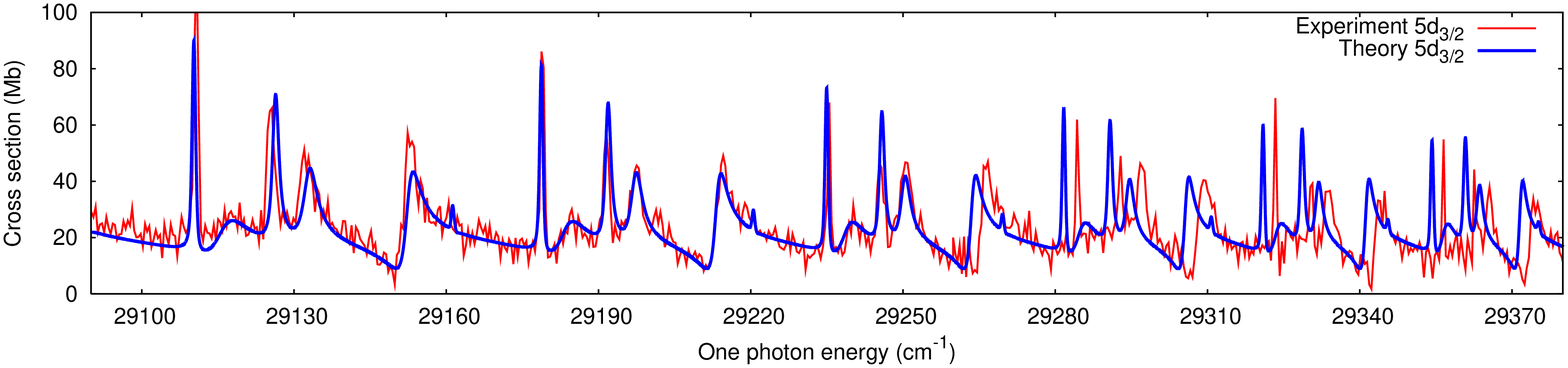}} \\
\resizebox{2.0\columnwidth}{!}{\includegraphics{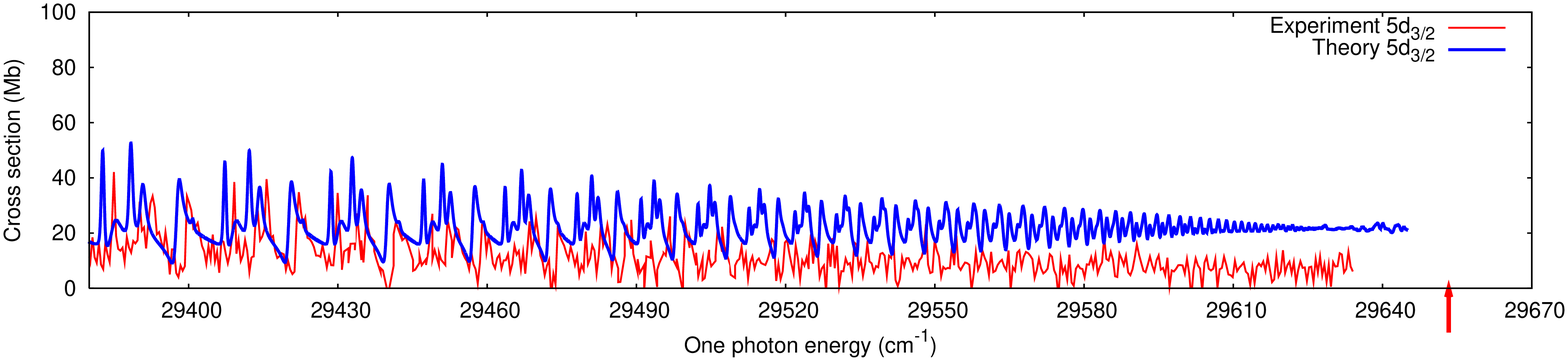}} \\
\end{tabular}
\end{center}
\caption{One photon cross sections: $5d_{3/2}$. \label{fig:AIRmany1d}}
\end{figure*}

\begin{figure}
\begin{center}
\begin{tabular}{cc}
\resizebox{0.5\columnwidth}{!}{\includegraphics*{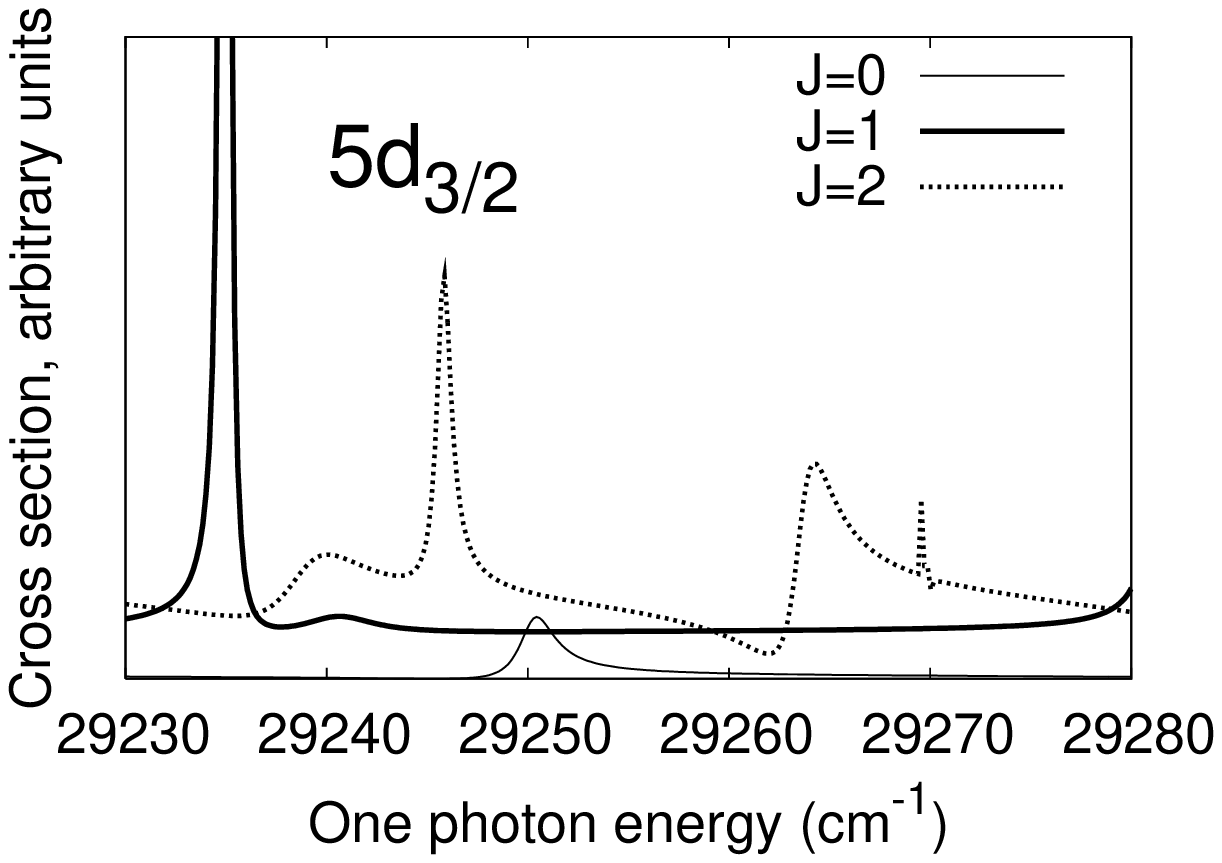}} &
\resizebox{0.5\columnwidth}{!}{\includegraphics*{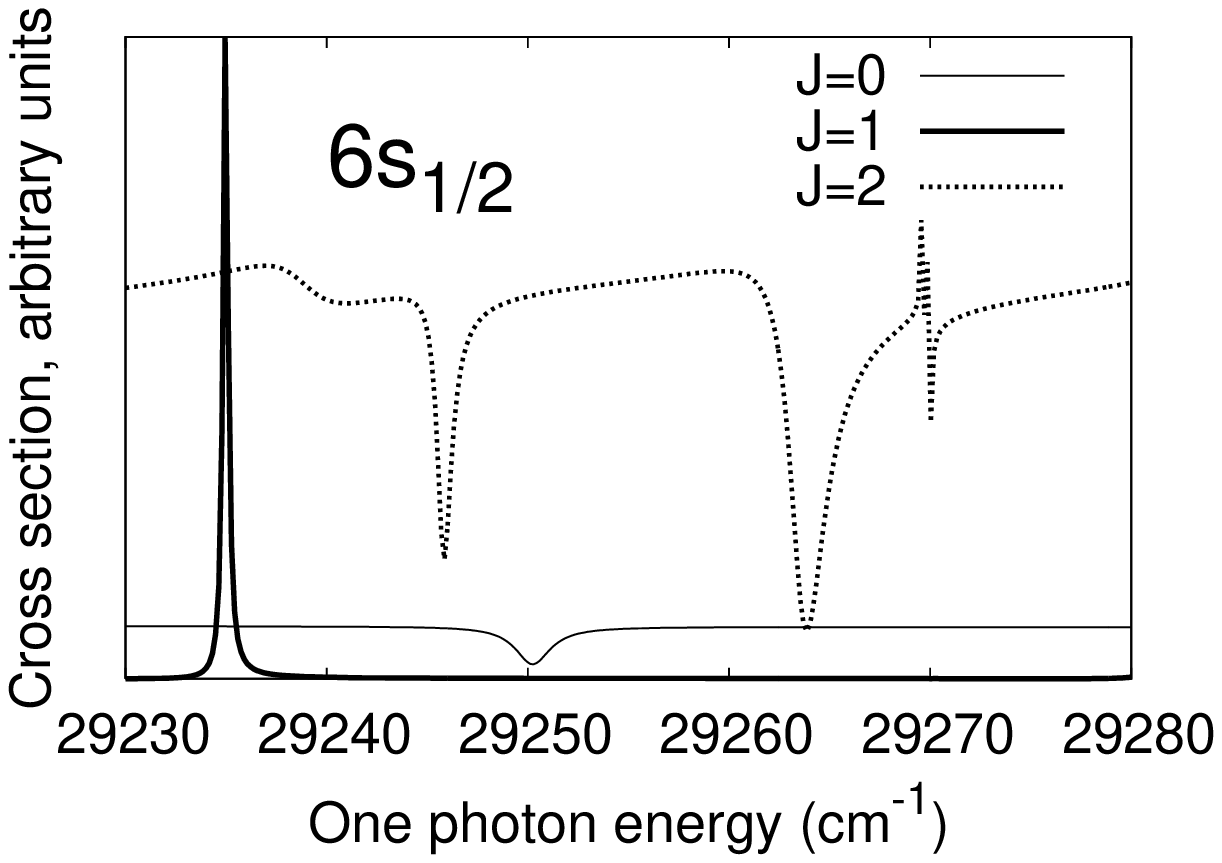}} \\
\end{tabular}
\end{center}
\caption{Zoom of one-photon isotropic~\cite{WGA93a} cross sections calculated separately for J=0, 1, and 2 final states. }\label{fig:peakzoom}
\end{figure}

Figs.~\ref{fig:AIRmany1s} and \ref{fig:AIRmany1d} depict
the one-photon partial cross section for the $6s_{1/2}$
and $5d_{3/2}$ channels.  There are several series of autoionizing resonances
visible in these results.  These can be identified by referring to Fig.~\ref{fig:peakzoom},
which shows cross sections calculated for one final angular momentum value $J$ at a time
over a small energy range.

Hyperfine depolarization is responsible for the presence of the $J=1$ final state resonance
series, and this series may be used to infer the amount of depolarization present in the
experiment.  The $J=1$ resonant autoionizing series is prominent in the 5$d_{3/2}$ final state cross section
but relatively smaller in the 6$s_{1/2}$ cross section.

We find that hyperfine depolarization effects~\cite{WGA93a} are insufficient to account for the
observed magnitude of the $J=1$ resonance series in the 5$d_{3/2}$ cross section.  The parameter
$g$ used in Ref.~\cite{WGA93a} indicates the amount of depolarization and for barium has a maximum
value of 1.0, corresponding to no depolarization, and a minimum of 0.896, corresponding to maximum
hyperfine depolarization.  We find that a value of $g=0.82$ produces the best agreement between
theory and experiment.  Therefore, it seems that there is depolarization in the experiment beyond
that describable via the hyperfine depolarization mechanism alone, which might reflect some combination of saturation and/or polarization impurity in the excitation lasers.

\subsection{Two-photon cross sections}

\begin{figure*}
\begin{center}
\begin{tabular}{c}
\resizebox{2.0\columnwidth}{!}{\includegraphics{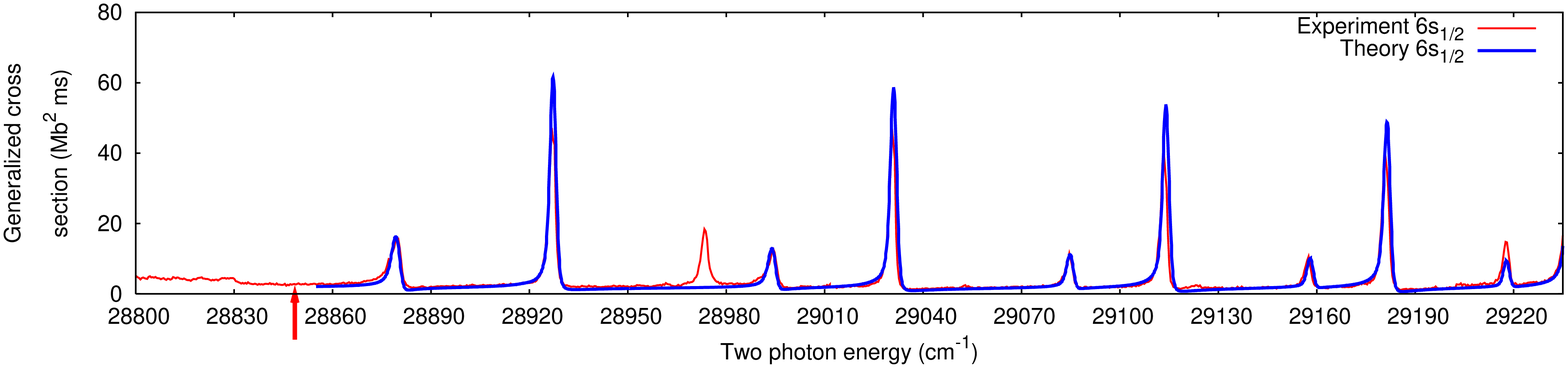}} \\
\resizebox{2.0\columnwidth}{!}{\includegraphics{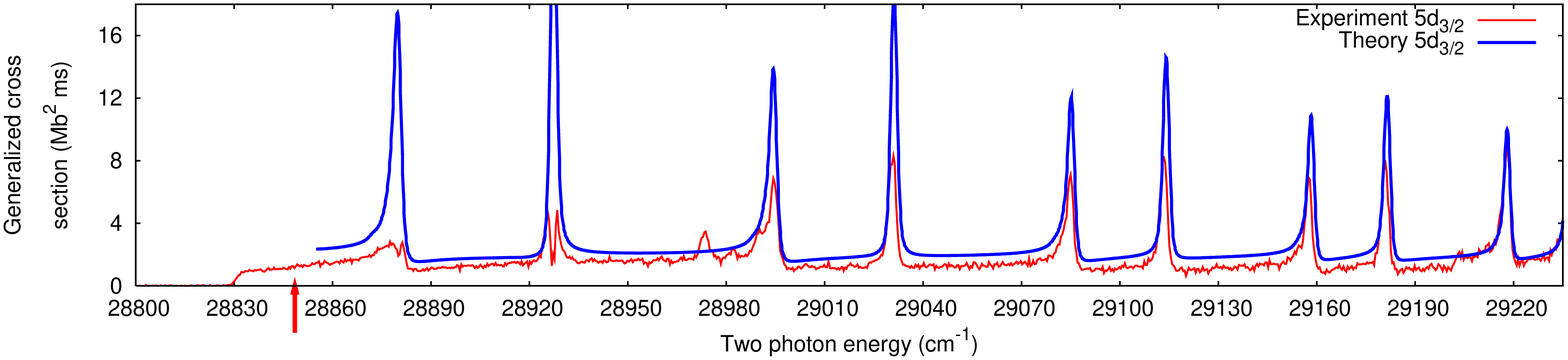}} \\
\end{tabular}
\end{center}
\caption{Two photon generalized cross sections: $6s_{1/2}$. \label{fig:AIRmany2s}}
\end{figure*}

\begin{figure*}
\begin{center}
\begin{tabular}{c}
\resizebox{2.0\columnwidth}{!}{\includegraphics{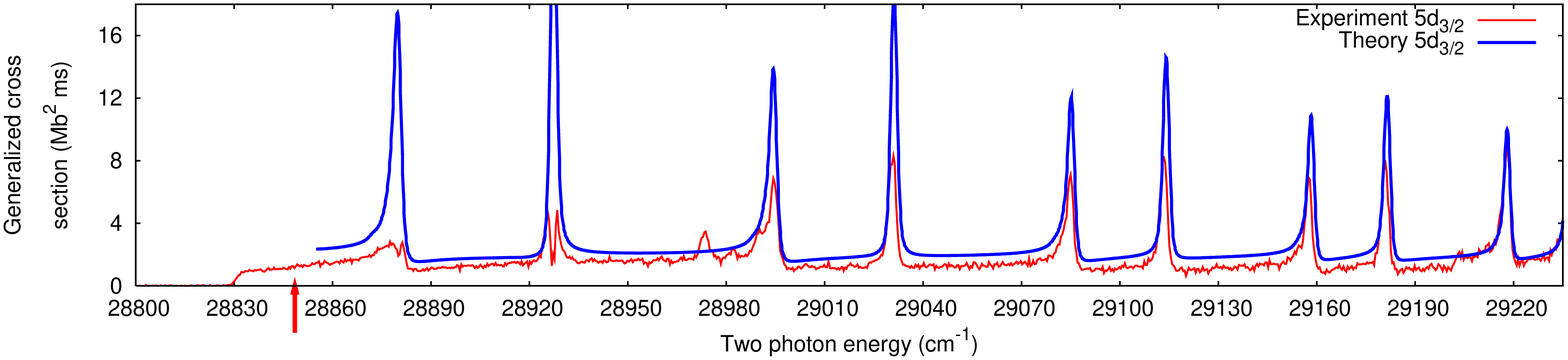}} \\
\resizebox{2.0\columnwidth}{!}{\includegraphics{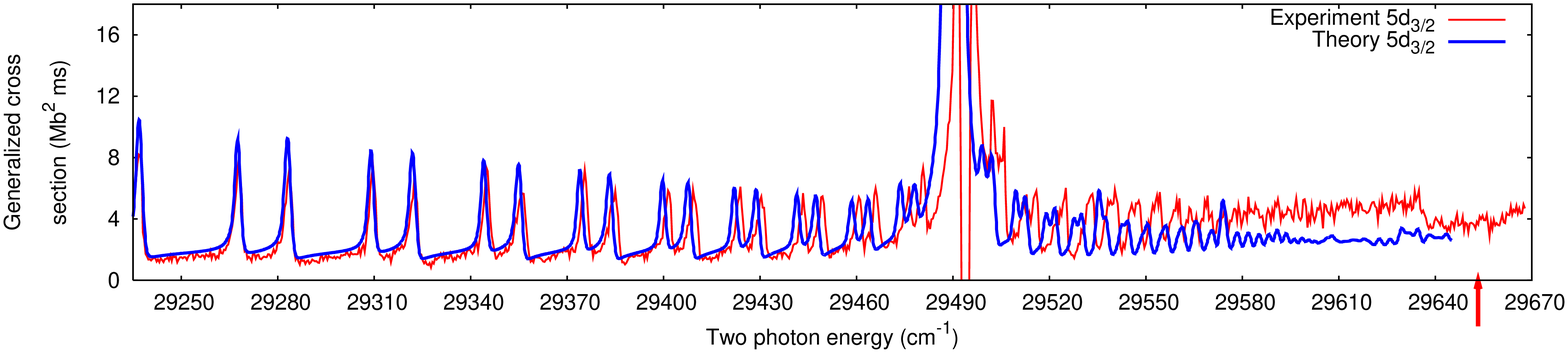}} \\
\end{tabular}
\end{center}
\caption{Two photon generalized cross sections: $5d_{3/2}$. \label{fig:AIRmany2d}}
\end{figure*}

Two-photon ionization cross sections are shown in Figs.~\ref{fig:AIRmany2s} and \ref{fig:AIRmany2d}.
The resonance series in the two-photon cross sections are members of one of two previously identified
autoionizing series,
$5d\ ^2D_{5/2} \ np_{3/2}$
and
$5d\ ^2D_{5/2} \ nf_{3/2}$~\cite{maeda}.
There is one experimental peak not reproduced by the theory, which lies at a two-photon energy of approximately
28974 cm$^{-1}$ (i.e. a red wavelength of 690.28 nm).  This energy corresponds to a one-photon transition to the $6s7p$ $^1$P$_1$ state, which lies at
32547.033 cm$^{-1}$ above the ground state, and it therefore seems that
this is an intermediate state resonance reached via a quadrupole transition from the initial state. We use the location of this peak to establish an absolute frequency calibration of our laser scan.  This peak in the two-photon spectrum allows us to calibrate our UV wavelengths as well, since these wavelengths are determined through measurements of the wavelength of the red laser beam before we double its frequency using the nonlinear crystal.  Therefore, we use this extra peak in the two-photon spectra to calibrate the {\it absolute} frequency shown in the measured spectra of Figs.~\ref{fig:AIRmany1s}, \ref{fig:AIRmany1d}, \ref{fig:AIRmany2s}, and \ref{fig:AIRmany2d}.

The strong peak around 678
nm (29489.5 cm$^{-1}$) in the red (two-photon) spectrum is due to the existence of an intermediate state resonance
(554 nm + 678 nm) with the $5d7s$ $^3D_1$ state, which lies 32805.169 cm$^{-1}$ above the neutral barium
ground state.
This intermediate state has total angular momentum $J=1$ and is therefore excited only to the extent that there is
depolarization of the laser-excited $^1$P$_1$ state (or else of impurity in the linearly polarized beam).  It allows us to obtain an
estimate of the amount of depolarization present in the experiment independent of the one-photon
results.  However, because of the fact that we have only one peak to which to refer, and because
the theoretical perturbation theory results are invalid in the immediate vicinity of the resonance
at 29489.5 cm$^{-1}$, the estimate of depolarization from the two-photon results is less precise than
is that from the one-photon results.  However, we find that a value of $g=0.82$ does produce better
agreement than does the value of $g=0.896$ corresponding to the maximum possible assuming only
hyperfine depolarization, corroborating the estimate obtained from the one-photon results that there
must be depolarization in the experiment beyond that expected from hyperfine depolarization alone.

We see evidence of the false counting at low electron kinetic energies that was
mentioned above, in the slow electron peaks in the red laser (two-photon)
spectrum in Fig.~\ref{fig:AIRmany2s} at 691.4 nm and 692.6 nm (28927 and 28877 cm$^{-1}$),
causing the double peak feature at these locations.

\begin{figure}
\begin{center}
\resizebox{1.0\columnwidth}{!}{\includegraphics*{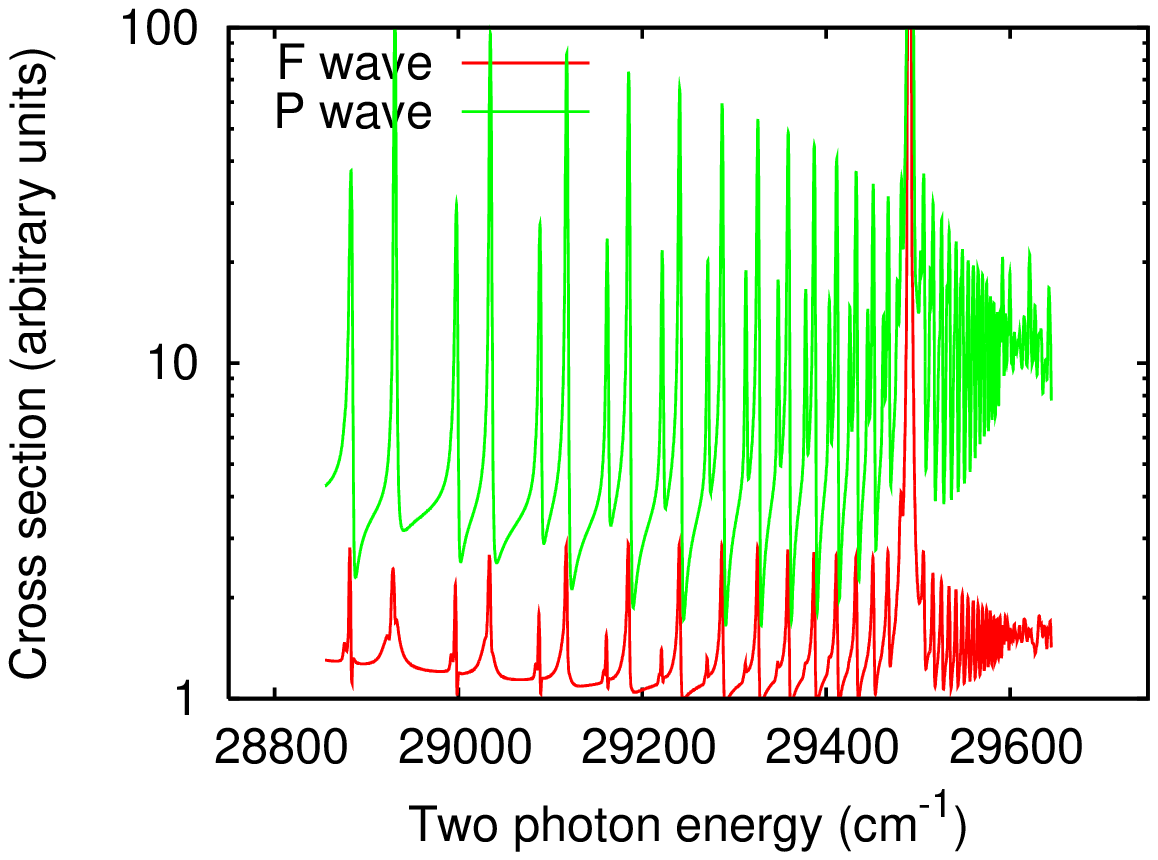}}
\end{center}
\caption{Cross sections (summed for both final electronic state channels)
as a function of the outgoing partial wave.\label{fig2}}
\end{figure}

We calculate that the ionized electron is mostly $p$-wave, with some $f$-wave, confirming the conclusions based upon photoelectron angular distributions measurements
to this effect in Ref.~\cite{YamazakiE07a,YamazakiE07b}, as can be seen in Fig.~\ref{fig2}.
The dominant final state total angular momentum for the two-photon ionization process is $J$=1.

The dominant intermediate state angular momentum is $J$=1 for the contribution from hyperfine
depolarization, and $J$=2 for the polarized contribution.  Of the J=2 intermediate virtual states that
contribute to the cross section, the dominant one is the 6$s$6$d$ $^1$D state, at 30236.828 cm$^{-1}$
(corresponding to a 2$\omega$ value of 24354 as plotted in the figures).  Also,
the 5$d$7$s$ $^3$D state at 32943.774 cm$^{-1}$ contributes to the cross section at the highest energy
range for the 5$d_{3/2}$ channel (above the intermediate state peak).

\section{Conclusions}

This combined theoretical and experimental study has given a quantitative understanding of the one-photon and two-photon photoionization spectra of barium, starting from the laser-excited $6s6p\ ^1P^o$ level.  This should now enable a careful study of the phase-dependent interference between one-photon and two-photon ionization of this barium excited state, and hopefully permit a quantitative test of this interference theory as well as the accuracy of the phases predicted by theory at the level described in the present study.

\section{Acknowledgments}

We thank J. P. D'Incao and S. T. Rittenhouse for helpful discussions.  This work was supported in part by the Department of Energy, Office of Science.  The material in the experimental section is based upon work supported by the National Science Foundation under Grant No. 0099477.

\bibliography{TwoPhoton}

\end{document}